# ATLAS Commander: an ATLAS production tool

V. Berten, L.Goossens
*CERN/EP/ATC, Geneva, Switzerland*

Chun L. Tan
*University of Birmingham, UK*

This paper describes the ATLAS production tool AtCom (for ATLAS Commander). The purpose of the tool is to automate as much as possible the task of a production manager: defining and submitting jobs in large quantities, following up upon their execution, scanning log files for known and unknown errors, updating the various ATLAS bookkeeping databases on successful completion of a job whilst cleaning up and resubmitting otherwise. The design of AtCom is modular, separating the generic basic job management functionality from the interactions with the various databases on the one hand, and the computing systems on the other hand. Given the near future reality of different flavors of computing systems (legacy and GRID) deployed concurrently at the various, or even a single ATLAS site, AtCom allows several of them to be used at the same time transparently.

## 1. INTRODUCTION

This paper describes the ATLAS production tool AtCom (for ATLAS Commander). The purpose of the tool is to automate as much as possible the task of a production manager: defining and submitting jobs in large quantities, following up upon their execution, scanning log files for known and unknown errors, updating the various ATLAS bookkeeping databases in case of success, cleaning up and resubmitting in case of failure.

The design of the tool is modular, separating the generic basic job management functionality from the interactions with the various databases on the one hand, and the computing systems on the other hand. How to interact with the various computing systems is defined separately in the form of plug-ins, which will be loaded dynamically at run time. The ability to access different flavors of computing systems, possibly even at the same time, in a transparent way, was an explicit design goal. At present time, many flavors of batch systems are in use at the different ATLAS institutes. Eventually, the GRID may simplify this situation, but at present different flavors of GRIDs exists as well, and in the near future it is very likely that GRID and non-GRID computing systems will be deployed concurrently, even at a single site.

The design of the tool assumes that jobs can be defined in a computing system neutral way. The current implementation features a virtual data [1] inspired approach equating job definitions with a reference to a transformation definition and actual values for its formal parameters. The transformation definitions include a reference to a script/executable, its needed execution environment in the form of 'used' packages, and a signature enumerating the formal parameters and their types.

AtCom is implemented in Java and can thus in principle run on any platform equipped with a Java run-time environment. It has been used on a regular basis from both Linux and Windows machines.

## 2. SHORT HISTORY

The design and development of AtCom started from scratch early September 2002 relying on the production experience gained by one of the authors in the months prior to the commencement of the AtCom project. Two months later, during the November ATLAS software week, a successful live demo was given submitting pile-up jobs to both the local CERN LSF cluster and the Nordugrid [2].

AtCom has been in continuous production use at CERN since October 2002. It has accounted for almost all production done at CERN. In total more than 10000 jobs spread over several hundreds of datasets (a dataset is a group of similar jobs) have been submitted.

The first production version, AtCom v1.0, was released end of January 2003, ending the period of official AtCom development and bringing the total resource count to about five man months. Subsequent unofficial development resulting in v1.2, released early March 2003, achieved the last remaining design goal: the ability to run the AtCom GUI part on a machine different than the one executing the commands using secure shell connections.

More information about the AtCom project can be found at [3]. The AtCom website contains links to a user guide, a developers guide, downloads, presentations, …

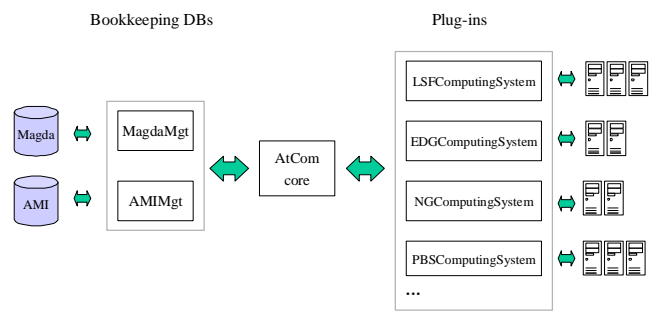

Figure 1: The AtCom architecture.

## 3. ARCHITECTURE

Figure 1 shows the top-level architecture of AtCom. In the middle is the AtCom core application that implements the logic of defining, submitting and monitoring jobs. It also holds the GUI that gives access to these functionalities. On the left are the two modules that interface AtCom to the ATLAS bookkeeping databases, respectively AMI (Atlas Meta-data Interface [4]) and Magda (Manager for Grid-based Data [5]). On the right there is the set of plug-ins that interface AtCom to the various flavors of computing systems. At present, the plug-in modules for the computing systems are loaded at run time, whereas the database modules are an integral part of the AtCom application. However, this is just an historical accident and an eventual next version of AtCom will also load the database modules at run-time as plug-ins that can be replaced easily.

The computing system plug-ins implement an abstract interface that defines methods and signatures for the usual operations: submitting a job, getting the status of a job, killing a job and getting the current output (stdout and stderr) of a job. A concrete plug-in is a Java class implementing this interface plus values for the additional configuration parameters it supports. For instance, the concrete plug-in LSF@CERN is based on the Java class LSFComputingSystem. The AtCom configuration file defines which plug-ins should be loaded, specifying for each, which Java class to use and what values for its configuration parameters to set.

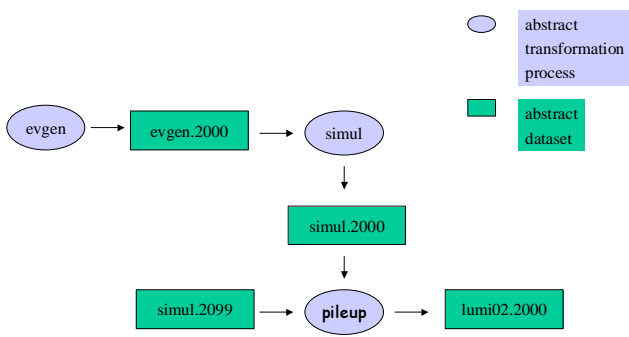

Figure 2: Datasets and abstract transformations.

## 4. UNDERLYING PRODUCTION MODEL

The underlying production model is based on the concepts of datasets, partitions, transformations and jobs. A dataset is a chunk of data that logically forms a single unit. An example is the dataset named dc1.002000.evgen that contains 15 million jet events. Because the total size of this dataset is close to 270 GB, it is for practical reasons split in 150 partitions each corresponding with a separate logical file. On the dataset level, abstract transformations create datasets based on a number of parameters and possibly taking one or more other datasets as input (figure 2). Again, for practical reasons, this transformation process is implemented using a number of concrete transformations, each coinciding with a single job operating on the partition level (figure 3).

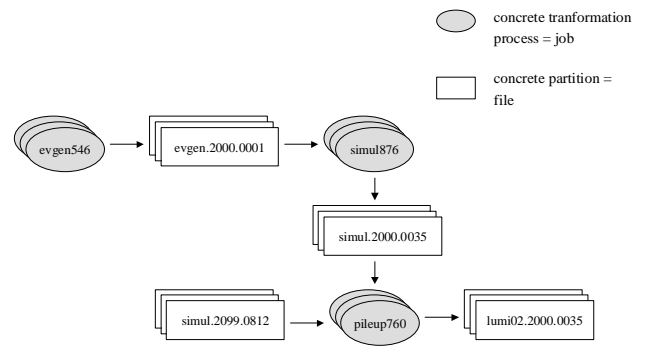

Figure 3: Partitions and concrete transformations.

## 5. BOOKKEEPING DATABASES

Within the bookkeeping information five logical database domains have been identified:
- physics meta-data database
- transformation catalog
- permanent production log
- transient production log
- replica catalog

The **physics meta-data database** holds information *about* the datasets: total number of contained events, physical properties like applied $p_t$ cuts, current status, etc. Most importantly, it records which abstract transformation was used to produce the dataset and what values were used for the formal parameters that are defined in the transformation's signature.

The **transformation catalog** records all existing transformations, identified by a name and a version. The simple versioning system (an additional string) allows to have a set of transformations that conceptually all do the same thing, but that differ e.g. in their genericity taking more or less parameters. At the same time, we also exploit it to indicate which transformations are supported in which release of the ATLAS software. Version strings are thus of the form <version>.<release>, e.g. "v4.602".

Transformation definitions contain the name of the executable (usually a script) that implements the transformation on the partition level, the signature of this executable, the physics signature, the output signature, the list of packages it depends upon, the name of the executable that will validate the job, and the name of the executable that will extract information from the job's stdout and stderr to update the permanent production log.

The signature is simply a list of formal names of all the parameters to be passed to the executable. However, the names can additionally be qualified with the keywords LFN or LFNlist. This means that the parameter is a file (list of files) subject to logical file name to physical file name translation. AtCom also supports the concept of a logical file name hint. This can be used to steer this translation process e.g. to only consider physical files stored in a particular place. The syntax for LFNlists supports simple enumeration, but there is also a shorthand for a range of files only differing in a single number.

The physicsSignature attribute lists the parameters on the dataset level. In practice this means that inputs are datasets instead of (logical/physical) file(s) and parameters only relevant on the file level are suppressed.

The output signature lists the formal names of the files that will be produced by the transformation.

The **permanent production log** records all partitions and all information relevant to them. Because partitions coincide with logical files they are identified by their logical file name (LFN). In the virtual data sense these are both the derivations and the invocations. They coincide because unlike in the virtual data approach, the data is computed only once.

The record holds all information necessary to produce the partition in a computing system independent way. It lists which transformation to use (a concrete version) and specifies the actual values for all the formal parameters. Additionally, it states for each formal output the mapping from the local file name (as produced by the executable) to a logical file name. Here as well a hint can be given to steer this translation.

The partitionStatus field will reflect the life cycle of the partition as it moves from 'defined' to 'submitted' and finally to one of 'validated', 'failed', 'undecided' or 'aborted'. Together with the reservedBy field it also allows to coordinate potentially different production systems and/or production users, ensuring that no partition is processed twice.

Finally, the records also hold information like the execution host, institute, file size, md5sum, number of events, etc. that can be filled in after the job has finished.

The **transient production log** is as well situated on the partition/job level, but records information that is only relevant while the job is running. Most notably, this includes some identification of the running job, i.e. identification of the computing system where it was submitted and a computing system dependent token to identify the job.

It should be noted that as this information is transient, the actual schema of this table is not crucial and different production systems may use entirely different ones. The one presented in this paper is merely the one used by AtCom.

Last but not least, there is the **replica catalog** that holds the mapping between logical files (partitions) and physical replicae of them all over the world.

Figure 4 shows the five logical database domains and how they are at the moment spread over the AMI and Magda database servers. Note that this is within the AtCom context. Within the larger ATLAS context other databases/servers hold similar information.

Figure 5 shows the different tables within AMI that are used by AtCom and their relations.

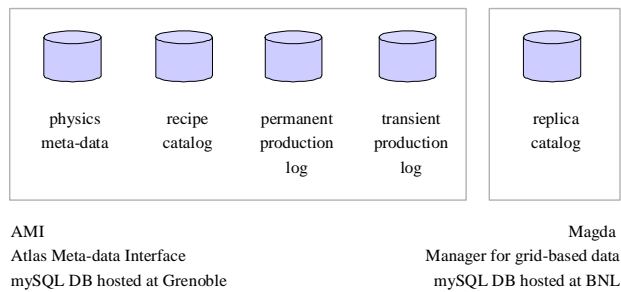

AMI                                                 Magda
Atlas Meta-data Interface            Manager for grid-based data
mySQL DB hosted at Grenoble      mySQL DB hosted at BNL

Figure 4: Logical database domains.

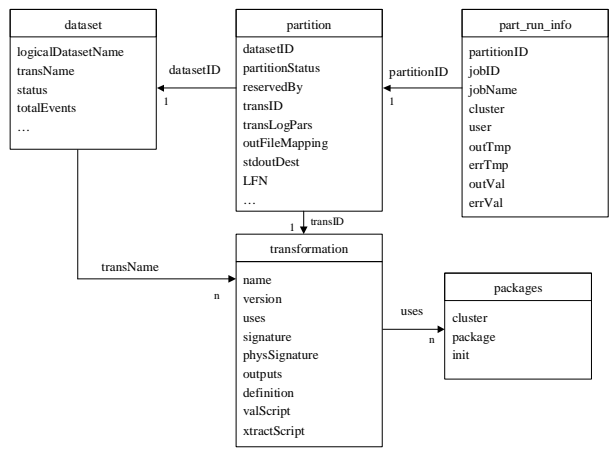

Figure 5: AMI tables used by AtCom.

## 6.  ATCOM FUNCTIONALITY

Basically, AtCom supports three classes of operations: job definition, job submission and job monitoring. These correspond with the three main panels of the GUI.

Figure 6: Partition creation panel.

## 6.1. Job definition

From the definition panel, the user can select a dataset she wants to define partitions with, by means of an SQL query composer (not shown but similar to the one shown in figure 7). She defines the fields of the dataset she wants to see and the selection criteria. Pull-down menus allow her to compose the most common queries, but the query text can, if needed, be arbitrarily edited. The search is executed and the result is displayed. She can then select a single dataset and then choose a particular version of the associated transformation. Based on this concrete transformation's signature AtCom will compose a form that will allow her to give values for all required parameters for all partitions she wants to define (figure 6).

The required parameters are a small number of 'constant' ones (i.e. they need to be given no matter what transformation is used e.g. the partition's LFN), the signature parameters, the output file mapping and the final destination for stdout and stderr.

In order to easily define a large number of partitions, AtCom allows her to define a range for the variable i and use its value in expressions for the parameters. The expression evaluator supports the usual basic arithmetic operations (+, -, *, /, %, parentheses) plus a way to format numeric results to a fixed number of digits (prefixing zeros if needed).

Additionally, the user can define a number of auxiliary variables (called A, B, C, …) and use them as well in the expressions. This allows one to capture and highlight similarities between the values of different parameters. Additionally, it allows one to capture as well the differences between similar datasets, facilitating the creation of a new dataset by re-using the definition of an existing one.

The amount of information that needs to be given is substantial. This is the price that needs to be paid for the genericity of both the transformations and the definition scheme in general. Fortunately, AtCom allows the user to save and retrieve the values she filled in (in AMI). In case of an attempt to retrieve the values for a dataset transformation combination that was not stored, it will return the best possible approximation available. Future versions of AtCom will probably feature a complementary simplified definition procedure that will exploit ATLAS conventions leaving fewer degrees of freedom and consequently fewer variables to be defined.

After having filled in all the parameters, the partition definitions can be previewed and, if satisfactory, they can be created.

## 6.2. Job submission

The second AtCom panel allows the user to submit any defined partition to any configured computing system. The user procedure is quite simple, what happens behind the scenes is the subject of section 7.1.

The procedure starts again with an SQL composer allowing you to retrieve a set of partitions (figure 7). The composer is slightly more sophisticated as it now allows showing and selecting upon attributes of both datasets and partitions.

Given a set of retrieved partitions the user can select an arbitrary subset and select a target computing system for submission. The jobs are submitted and automatically transferred to the next panel for monitoring.

Partitions can also be transferred to the monitoring panel without submission. This is perhaps a bit strange, but it allows the exploitation of some of the features of the monitoring panel, most notable the ability to modify the status of the partitions directly. The monitoring panel also shows a bar/pie chart plotting the number of monitored partitions in each state. So to get an overview of a complete dataset one may need to artificially move the 'defined', 'validated' and 'aborted' partitions to the monitoring panel as there is normally no record associated with them in the transient production log database.

In section 7 we will in detail present what happens when one submits a job.

Figure 7: SQL composer.

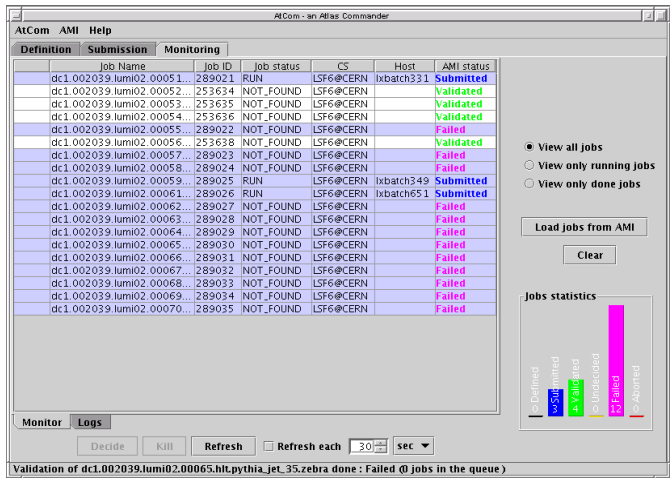

Figure 8: Monitoring panel.

## 6.3. Job monitoring

The monitoring panel (figure 8) shows for each job being monitored
- job name
- job ID
- status
- computing system (CS)
- host
- AMI status

Job name is the name given to the job by AtCom. By convention it is simply the LFN of the partition. Job ID is the computing system dependent token to identify the job. Status is the status as reported by the computing system. Possible values are 'running', 'wait', 'done', 'error', and 'failed'. AMI status is the status of the partition as stored in the permanent production log database.

The panel allows the user to check the status of all monitored jobs on demand, or poll automatically at regular intervals. Additionally, the user can select a number of jobs and right click on them to invoke one of a large set of operations:
- kill
- decide
- refresh
- submit
- resubmit
- show
- revalidate
- set AMI status

'Kill' will issue the command to stop execution of the job. 'Decide' allows the user to open up the decision dialog (more on this shortly). 'Refresh' triggers the status polling only for the selected jobs. 'Submit' submits a partition in status 'defined' to one of the computing systems (identical to submission panel function), while 'resubmit' will submit the job again to the same computing system it ran on before (only if status is

**MONT002**

'failed'). 'Show' gives access to the various files that are associated with the job at hand (including the stdout, stderr if the computing system supports such functionality) and the full status text as reported by the computing system. 'Revalidate' re-initiates the validation procedure (to be described shortly). 'Set AMI status', allows the AMI status of partitions to be directly manipulated. Note however that only a very limited subset of state transitions can sensibly be initiated by the user. Requesting any other will result in a warning and will have no further effect.

When a job moves from 'running' to 'done', post-processing is automatically started. The validation script name is resolved into the name of a real executable and it is called with the job's stdout and stderr files as parameters. By convention these validation executables must return 1 if the job executed successfully, 2 if the success/failure cannot be decided automatically (e.g. the string 'error' occurs in the log but not in a known way), or 3 if the job failed in a known way.

If the job is ok, the output files are registered with the replica catalog (Magda). The extract script is resolved into an executable and run over stdout and stderr. By convention the extract scripts are supposed to write on stdout a set of attribute value pairs that will be used by AtCom to attempt an update query on the partition record of the job at hand. Logfiles are copied to their final destination (as defined in the partition record) and finally the partition's status is set to 'validated'.

If the job failed, the outputs as defined in the partition's output mapping are deleted and the status is set to 'failed'. If the job is 'undecided', the status is changed accordingly, pending a decision by the user. This decision can be made conveniently through the decision dialog (figure 9). It will show for the job at hand the output of the validation script and the job's stdout/stderr.

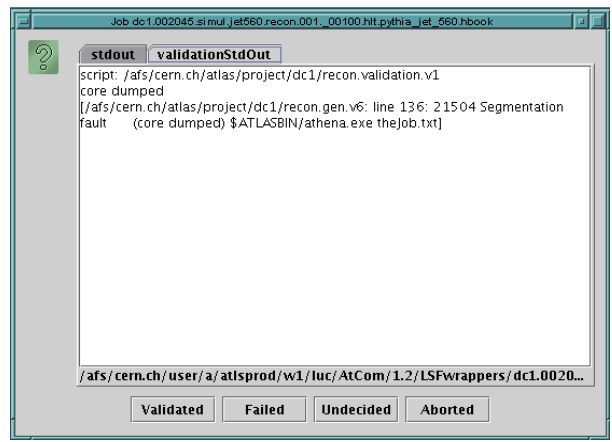

Figure 9: Decision dialog.

## 7. COMPUTING SYSTEM PLUG-INS

The purpose of a computing system plug-in is to present the different flavors of computing systems in a uniform way to AtCom. This uniform interface is very simple: a small number of methods (submit a job, kill a job, get status of the job, get current stdout/stderr of job) and a mapping between the possible states of the job within each computing system and the 'abstract' states as defined by AtCom.

In total five plug-ins have been developed. In the following subsection we first examine in detail the LSF plug-in. A second subsection lists all remaining plug-ins and their current status.

It is our experience that development of a new plug-in generally takes only one person-day.

## 7.1. LSF plug-in

In this section we describe the workings of the LSF plug-in. Together with the EDG plug-in it was one of the first two plug-ins developed, and effort was made to make its workings as much grid-like as possible. Nevertheless, for practical reasons, it also exploits a lot of features unique to the only place were it is currently deployed: CERN. It also served as a starting point for other non-grid plug-ins like the one for PBS and BQS.

Almost all of the LSF plug-in code is concerned with the procedure of submitting a job, as described below.

First of all, an attempt is made to reserve the partition in the permanent production log database. If this succeeds, a record is created in the transient production log database. Then follows the bulk of the work: creating the actual wrapper script that will be submitted.

The wrapper starts with the commands needed to set-up the correct environment. To this end, for each used package the code to be inserted is looked up in the package database using the key <packageName, computingSystem>, e.g. <Atlas603,LSF@CERN>.

Next, all logical actual parameter values are converted into their physical counterparts. For LFN parameters a Castor (CERN's hierarchical storage system) path is prefixed according to a fixed algorithm, exploiting the adopted conventions of organizing the castor directory tree. The algorithm also takes into account the LFN hint, if present. For LFNlist parameters first the LFN ranges are expanded, and then each LFN is converted as described above. Finally, code is inserted into the wrapper writing all these logical file names into an auxiliary file whose name will be passed as value to the core transformation executable.

Note that in general the plug-in should consult the replica database (e.g. through the MagdaMgt module in the case of AtCom) to do this translation. The current approach avoids that we would not only have a CERN specific LSF plug-in, but also a CERN specific replica module.

Next, a line is inserted into the wrapper calling the core executable using the converted actual values.

Finally, for each declared output of the transformation code is inserted to move the output to its final destination. This as well involves a translation of the logical file names into physical castor file names based on the adopted conventions and possibly exploiting the LFN hint.

Note that the outputs are not registered with the replica catalog by the job itself. An alternative strategy, adopted e.g. by the EDG plug-in, would be to copy and register the file from the job. Registering the files only after they have been validated has an obvious advantage. Additionally, it avoids a wide area network dependency.

The wrapper code is saved with a unique name in a unique job directory within the LSF plug-in working directory. The job is submitted using the command from the configuration file (allowing you to specify additional options) using the –o and –e options to channel the stdout and stderr to temporary files in the job directory. The jobID returned is stored in the transient production log record together with the temporary locations of the stdout and stderr.

## 7.2. Other plug-ins

Besides the LSF plug-in, which has been extensively used and is continuously enhanced, there is a small number of other plug-ins in various states of usage and maintenance.

- EDG (European Data Grid) plug-in

has been maintained actively since the beginning of the AtCom project but has not been used for any real production; will be deployed during summer 2003 on the LCG-1 production facility;

- NG (Nordugrid) plug-in

was developed as a replacement for the EDG plug-in for the life demonstration in November 2002; after that development stopped due to lack of interest from the Nordugrid side;

- PBS (Portable Batch System) plug-in

developed more recently; operational and maintained but not (yet) used in productions;

- BQS (Batch Queuing System) plug-in

developed more recently; BQS is the batch system deployed at CCIN2P3 (Lyon).

## 8. CONCLUSIONS AND OUTLOOK

As stated in the introduction, AtCom is a tool for production managers. It is a convenient but rather thin layer on top of both the bookkeeping databases and the computing systems. We do not think that in its current form it is suitable for use by a broader audience. The protection against doing things wrong is just too minimal. Additionally, it is not a convenient tool to follow up the execution of *thousands* of jobs. For this its interactiveness is prohibitive rather than convenient.

AtCom has been used extensively at CERN now for more than half a year and consequently has become optimally fit for the specific type of productions that take place there. CERN usually is the first ATLAS site to run any ATLAS code in production mode and consequently often discovers possible error conditions while running. Pre-productions are started, closely monitored, aborted and restarted, etc. Additionally, it has become customary that CERN processes the many smaller datasets, while outside institutes process a few smaller datasets or even just part of a single bigger dataset. Of the 541 datasets currently registered in meta-data catalog, more than 410 were processed completely at CERN. The remaining 130 are spread over about 40 sites. This is a clear indication of the unique role of CERN within ATLAS.

When dealing with a few hundred partitions of a single or a few datasets, it is possible to survive without tools. Conversely, when you need to worry about thousands of jobs of a single or a few datasets you are probably better of with a non-interactive tool suite (like the GRAT [6] system deployed on the USGRID).

Even though AtCom's user base has been extremely small, it has been a major driving force in defining the bookkeeping databases, has acted as a catalyst for defining an ATLAS-wide uniform production framework (to be gradually introduced in the course of 2003), and has made a substantial contribution to this framework.

The coming months will be decisive for the future of the ATLAS production strategy. The uniform production framework will become reality. ATLAS datasets, partitions and transformations will be stored in a single logical database. The non-automated production mode, involving many people all over the world, will gradually be phased out because of the high risk of human errors. Highly automated production tools will take care of almost all the production at all possible sites. The production model will be extended to take into account productions on the scale of physics groups all the way down to the scale of a single physicist. Complementary tool suites targeted more at these new audiences (e.g. GANGA [7], DIAL [8]) will be integrated and deployed.

For AtCom the most likely scenario is that it will continue to exist, serving the very specific needs of the CERN production team. We expect that at some point the need for different plug-ins will be made obsolete by GRID technology, presenting all computing resources through a uniform interface.

Another possible scenario is that AtCom will be deployed as a GUI upon the highly automated production system, which at the moment has no GUI at all. This way one would be able to seamlessly switch from interactive to automatic production mode and vice versa.

## Acknowledgments

The AtCom project would not have been possible without the CERN technical student program funding its main designer and programmer.

On the plug-in side we would like to thank the EDG team, and especially J.J. Blaising, for their kind assistance with creating and testing the EDG plug-in. Thanks to the Nordugrid team for doing the same for the NG plug-in. Thanks to J. Fulachier for co-developing the BQS plug-in. Finally, we would like to thank GridPP(UK) for supporting the creation of the PBS plug-in, and special thanks to P. Watkins, P. Faulkner and L. Lowe from the University of Birmingham for their technical assistance and support.